# Monte Carlo Framework for Physics-Based Design and Optimization of X-Ray Compressive Ghost Imaging Systems

R. H. Shukrun,[1] A. Ben-Yehuda,[1] Y. Korotinsky,[2] Y. Klein,[1] O. Sefi,[1] Y. Fried,[2] L. Epstein,[2] and S. Shwartz[1,*]

[1] Physics Department and Institute of Nanotechnology, Bar-Ilan University, Ramat Gan 52900, Israel

[2] Radiation Safety Department, Soreq Nuclear Research Center, Yavne 81800, Israel

[*] sharon.shwartz@biu.ac.il

**Abstract:** X-ray ghost imaging has emerged as a promising computational imaging modality capable of forming images from spatially integrated measurements. Unlike optical implementations, image formation is governed by multiple energy-dependent photon-matter interactions: photoelectric absorption, Compton scattering, and fluorescence. The relative contributions of these interactions depend on the source spectrum, sample geometry and composition, and detector configuration. We present a FLUKA-based Monte Carlo framework that models photon transport and predicts bucket signals generated by binary illumination patterns. The framework couples a physical scaling analysis, which converts the simulated bucket signals into photon numbers, tube loading and absorbed dose, with a ghost imaging reconstruction analysis of image quality and convergence. Our framework helps assess whether a given configuration is limited by the source output, the mask modulation or the reconstruction before any measurement is made. We evaluate the reconstruction for a centimeter-scale imaging system for an iron crown using one transmission detector and three off-axis scattering detectors over a range of source spectra. The results reveal how source characteristics and detector geometry influence reconstruction quality and radiation dose, providing quantitative guidelines for future X-ray ghost imaging system design. More broadly, this work establishes a predictive design framework for application-specific X-ray ghost imaging systems.

## Introduction

Computational ghost imaging (CGI) is an imaging modality in which an object is illuminated by a sequence of known structured illumination patterns, while the transmitted or scattered radiation is measured by a non-spatially resolving (bucket) detector. The image is then reconstructed computationally from the measured bucket signals and the corresponding illumination patterns [1,2]. When combined with compressive sensing, CGI can recover high-quality images from substantially fewer measurements than required by conventional sampling, improving acquisition efficiency [2,3].

Extending computational ghost imaging to the X-ray regime combines the high penetration power, material sensitivity, and elemental contrast of X-rays with the flexibility of computational image formation [4,5]. This combination has enabled a wide range of ghost-imaging-based modalities, including ghost tomography, ghost phase-contrast imaging, energy-selective ghost imaging, fluorescence-based elemental ghost mapping, and time resolved ghost imaging using synchrotron, tabletop, and laboratory X-ray sources [6–11]. More recently, ultra-low-light megapixel X-ray ghost imaging based on a synthetic-aperture architecture further demonstrated the potential of computational measurement strategies for improving photon efficiency while maintaining high spatial resolution [12]. In parallel, X-ray CGI has been extended to human-organ-scale fluorescence imaging, illustrating the growing versatility and practical applicability of computational X-ray imaging [13]. Recent developments have further expanded the practical scope of X-ray CGI through magnified imaging with polycapillary optics, rapid acquisition using continuously scanned aggregate patterns, full-field coded-aperture X-ray fluorescence imaging, and highly compressive volumetric X-ray fluorescence ghost tomography [14–17].

However, designing X-ray CGI systems is particularly challenging because the measured bucket signal is governed by the combined, energy-dependent contributions of multiple photon-matter interactions. Since the bucket detector records an integrated response, these contributions are superimposed and are not readily separated without additional discriminating information. This complexity is amplified in laboratory X-ray systems, where broad polychromatic spectra span energy ranges over which the relative contributions of these interaction mechanisms vary continuously. As illustrated in Fig. 1, this energy dependence makes the source spectrum, imaging modality (transmission, scattering, or fluorescence), object geometry, detector configuration, and illumination strategy strongly coupled design parameters [18]. Capturing these coupled effects accurately is difficult with simplified analytical forward models, motivating the use of physics-based radiation-transport simulations for predictive system design. These developments have considerably expanded the range of X-ray CGI architectures and applications, increasing the need for predictive tools to guide system design and optimization [11].

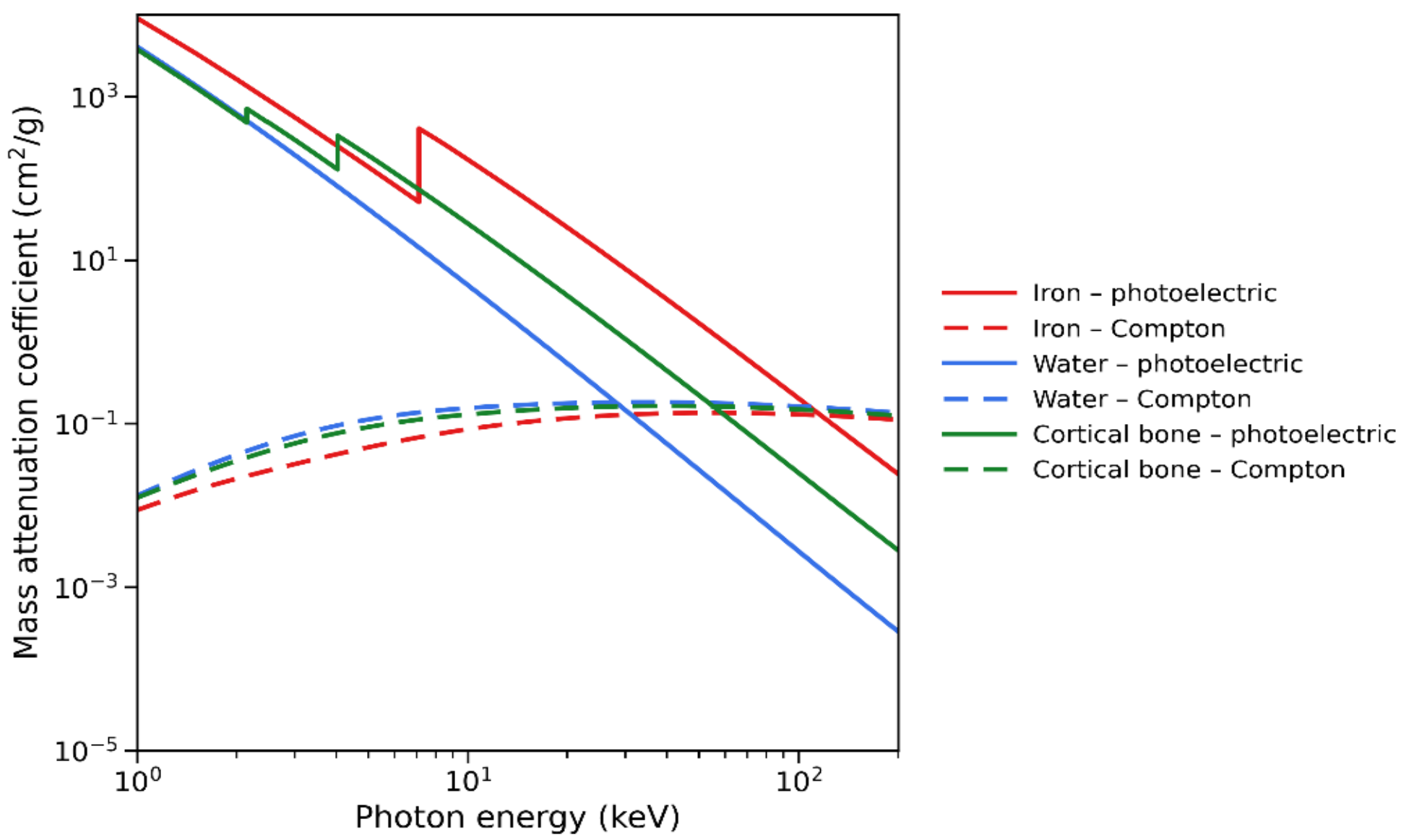

**Fig. 1.** Energy dependence of representative X-ray interaction mechanisms. Mass attenuation coefficients for photoelectric absorption and incoherent Compton scattering in iron, liquid water, and cortical bone [18]. Red, blue, and green curves correspond to iron, water, and cortical bone, respectively; solid lines denote photoelectric absorption and dashed lines denote Compton scattering. The variation of the interaction strengths with photon energy and material composition illustrates the spectrum dependence of the imaging physics considered in the proposed design framework.

Here, we present a predictive framework based on Monte Carlo radiation-transport simulations for the physics-based design and optimization of practical X-ray CGI systems. Monte Carlo radiation-transport simulations provide a high-fidelity approach for modeling X-ray photon transport in realistic imaging configurations, accounting for complex object geometries, and energy-dependent interaction physics [19,20]. The framework enables quantitative evaluation of alternative source spectra, detector configurations, acquisition strategies, and imaging geometries before experimental implementation. By simultaneously predicting imaging performance and absorbed dose, it provides a practical tool for optimizing X-ray CGI systems under the constraints imposed by ionizing radiation. Previous studies by our group have applied Monte Carlo simulations to scattered-photon computed tomography and X-ray fluorescence ghost imaging [13,21]. Here, we integrate photon-transport modeling, bucket-signal statistics, compressive reconstruction, and absorbed-dose estimation within a common workflow for comparing X-ray CGI configurations.

The same simulated illumination ensemble supports the evaluation of channel-specific bucket signals, reconstruction behavior, and ideal photon-counting requirements, while distinguishing Monte Carlo statistical uncertainty from physical counting noise. We demonstrate this workflow by comparing source spectra and individual and combined detection channels for a complex three-dimensional iron object.

**Results**

In CGI, the object is sequentially illuminated by a set of known structured illumination patterns, $I_i(r),\ i = 1, \dots, N$ [1,2]. For each illumination pattern, the transmitted, scattered, or fluorescent radiation is collected by a non-spatially resolving (bucket) detector, producing a single integrated measurement. Collecting the measurements for all illumination patterns yields the measurement vector $b$, which can be expressed as the linear inverse problem [2,3].

$$b = Ax + \eta. \tag{1}$$

Here $A$ is the matrix whose rows correspond to the illumination patterns, $x$ is the unknown object response, and $\eta$ accounts for measurement noise. For an image containing $P$ pixels, conventional CGI reconstruction generally requires a comparable number of measurements [2]. However, when suitable prior information, such as sparsity or low total variation, is available, compressive CGI enables image reconstruction using fewer measurements than the number of image pixels [2,3].

The linear measurement model in equation (1) is independent of the physical origin of the bucket measurements. The central challenge addressed in this work is therefore not the reconstruction itself, but the accurate prediction of the measurement vector b for realistic X-ray imaging systems, where each bucket measurement results from complex photon transport through the object and imaging system.

To address this challenge, we developed a predictive framework for the design and quantitative analysis of X-ray compressive CGI systems, implemented using FLUKA-based Monte Carlo radiation-transport simulations [19,20,22,23]. The framework predicts the bucket measurements associated with each illumination pattern by explicitly modeling photon transport through the complete imaging system. FLUKA provides a comprehensive and extensively validated platform for modeling X-ray transport in realistic imaging configurations, including complex object geometries and energy-dependent interaction physics. The workflow of the proposed framework is illustrated schematically in Fig. 2.

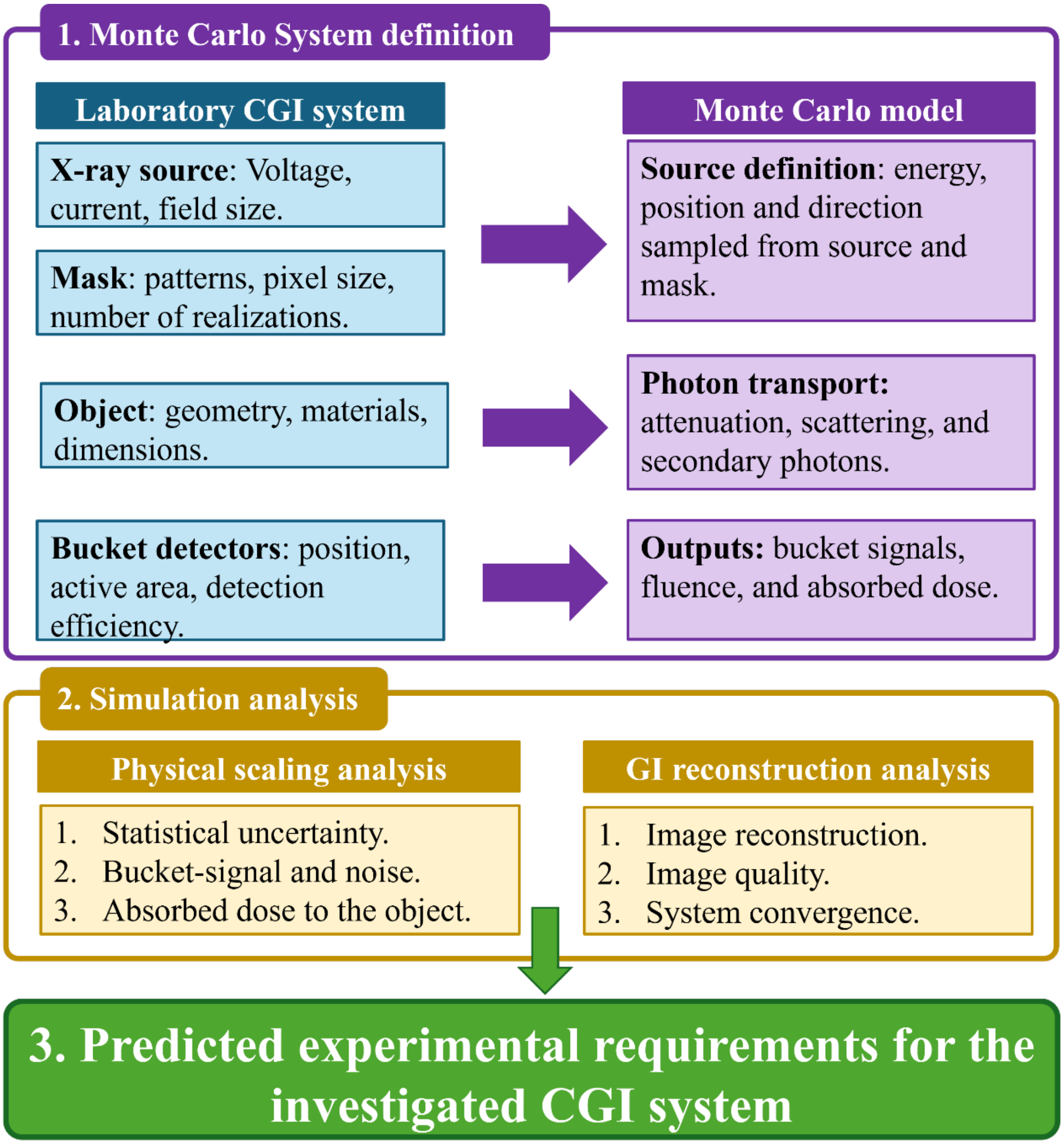

**Fig. 2.** Workflow of the proposed Monte Carlo framework for the predictive design of X-ray CGI. The laboratory imaging configuration, including the source, illumination mask, object, and detector arrangement, is translated into a Monte Carlo radiation-transport model, as described in the Methods. The simulated bucket signals are subsequently analyzed in two parallel analyses: a physical scaling analysis, which evaluates the statistical uncertainty, the bucket signal and its noise, and the absorbed dose to the object, and a CGI reconstruction analysis, which evaluates image reconstruction, image quality, and convergence with the number of realizations. Together, these analyses are used to evaluate the performance of candidate imaging configurations and estimate the experimental requirements for achieving the desired imaging performance.

We demonstrate the framework by analyzing a 3-centimeter-scale iron crown shown in Fig. 3a. The crown mesh model was retrieved from a publicly available online platform [24]. Its complex geometry and energy-dependent interaction properties provide a representative benchmark for evaluating practical X-ray CGI systems. The imaging system, schematically illustrated in Fig. 3a, includes one transmission detector and three off-axis scattering detectors (referred here as Up, Left and Right). The full detector setup is defined explicitly in the Methods, including detector positions, orientations, scoring boundaries, and the reference-scoring volume. Simulations were performed for three source spectra - 60 kVp, 150 kVp, and hardened 150 kVp, as shown in Fig. 3b [25]. The same ensemble of 4,000 random binary illumination patterns was used for all three source spectra, with each pixel independently assigned as open with a probability of 0.5. Each realization was simulated with $N_p = 5 \cdot 10^7$ primary photons, giving an estimated relative standard deviation of approximately 0.9% in the photon count per open mask pixel, as described in the methods. Scattered-photon imaging provides a particularly demanding test of the framework because the bucket detector signal depends strongly on the source spectrum, attenuation, multiple interactions, and detector geometry**.** We therefore use this configuration to examine how the source spectrum and detector geometry affect reconstruction quality, convergence with the number of illumination patterns, and absorbed radiation dose.

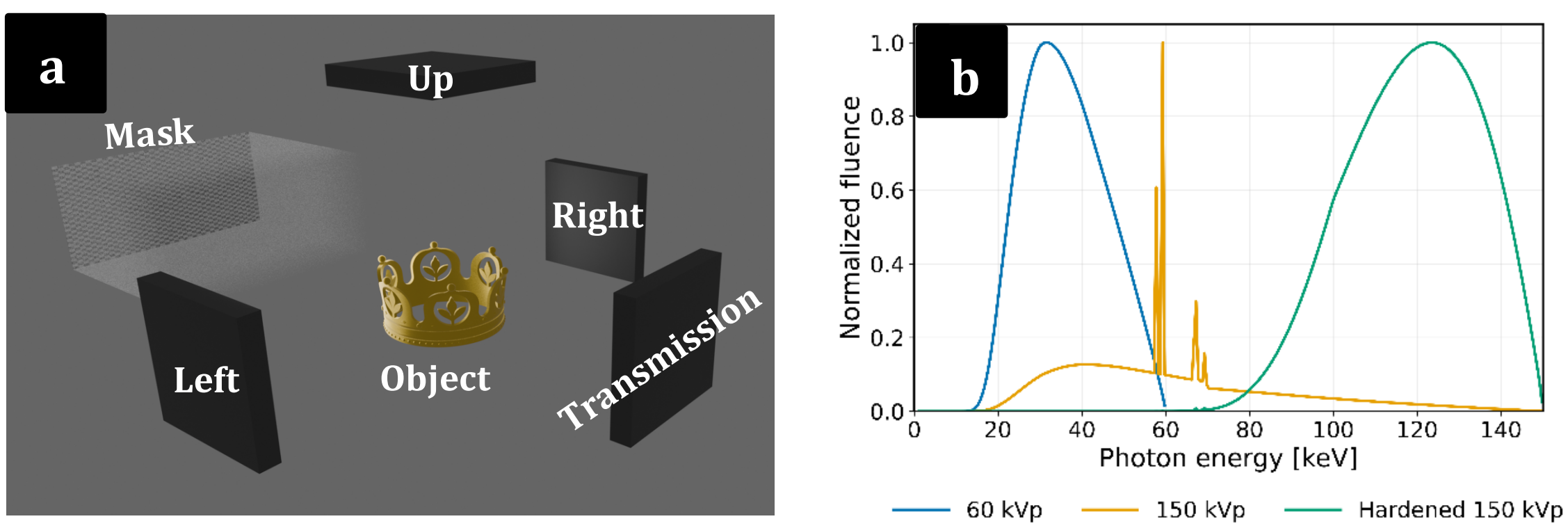

**Fig. 3.** Simulated X-ray CGI system and source spectra. (a) Schematic representation of the simulation geometry, showing the structured-illumination mask, the 3-cm-diameter iron crown, one transmission detector, and three off-axis scattering detectors denoted Up, Left, and Right [24]. (b) Source spectra used in the simulations: 60 kVp (blue), 150 kVp (orange), and hardened 150 kVp (green), each normalized to its own maximum spectral fluence for display. These spectra were used to evaluate the influence of source energy distribution and detector geometry on CGI reconstruction performance, convergence, and absorbed dose [25].

To relate the Monte Carlo bucket signals to the design of a physical experiment and to estimate the source fluence required per realization, it is necessary to establish what level of bucket-signal fluctuations relative to measurement noise is sufficient for image reconstruction. Klein et al.

investigated how reconstruction quality depends on the magnitude of the bucket-signal fluctuations relative to noise and on the compression ratio [5]. For this purpose, they defined a fluctuation-to-noise ratio (FNR) as the standard deviation of the measured bucket intensities across the illumination realizations divided by the corresponding noise magnitude:

$$\mathrm{FNR} = \frac{\sigma_s}{\sigma_n}, \tag{2}$$

where $\sigma_s$ is the standard deviation of the bucket intensities s across the illumination realizations, and $\sigma_n$ is the corresponding measurement-noise magnitude. In the analysis of Klein et al., the measurement noise was assumed to be dominated by Poisson shot noise. Substituting the corresponding noise magnitude into equation (2) gives [5]:

$$\mathrm{FNR_P} = \frac{\mathrm{std}(s)}{\langle\sqrt{s}\rangle}, \tag{3}$$

where $\mathrm{FNR_P}$ denotes the fluctuation-to-noise ratio obtained under the Poisson shot-noise assumption, and $\langle\cdot\rangle$ denotes the average over all illumination realizations. Klein et al. showed that, for the configurations investigated in their study, no clear image was obtained for $\mathrm{FNR_P} < 1$, reconstruction quality improved with increasing FNR, and further increases beyond approximately $\mathrm{FNR_P} = 5$ produced little additional improvement in image quality [5]. Motivated by these findings, we adopt $\mathrm{FNR_P}=5$ as a representative reference value for comparing the ideal photon-counting requirements of the configurations investigated here. When applying this definition to the simulated data, the bucket-signal standard deviation was represented by the Monte Carlo-noise-corrected mask-induced standard deviation, $\sigma_{\mathrm{mask}}$, defined in the Methods.

Before examining the reconstructed images, we first determine whether the Monte Carlo calculation resolves the mask-to-mask bucket variations that carry the CGI information. To quantify this, we introduce the Monte Carlo fluctuation-to-noise ratio ($\mathrm{FNR_{MC}}$), defined as

$$FNR_{\mathrm{MC}} = \frac{\sigma_{\mathrm{mask}}}{\sigma_{MC}}, \qquad \sigma_{\mathrm{mask}} = \sqrt{\sigma_{obs}^2 - \sigma_{MC}^2}, \qquad \sigma_{MC} = \sqrt{\frac{1}{N}\sum_{i=1}^{N}\sigma_i^2}, \tag{4}$$

where $\sigma_{obs}$ is the observed standard deviation of the simulated bucket signals across illumination realizations, $\sigma_{MC}$ is the root-mean-square Monte Carlo statistical uncertainty over the $N = 4000$ realizations, and $\sigma_i$ is the absolute Monte Carlo statistical uncertainty of the bucket signal for realization i, as defined in the Methods. Here, $\sigma_{\mathrm{mask}}$ is the Monte Carlo-noise-corrected mask-induced standard deviation, calculated by subtracting the Monte Carlo statistical variance from the observed bucket-signal variance and taking the square root, as detailed in Eqs. (8) and (9) in the Methods, whereas $\sigma_{MC}$ represents the typical finite-history statistical uncertainty of an individual processed $N_p = 5\times10^7$ primary realization. $FNR_{\mathrm{MC}}$ characterizes the numerical precision with which the bucket modulation is resolved by the Monte Carlo calculation, whereas $FNR_{\mathrm{P}}$ characterizes the photon statistics required for that modulation to be resolved in an ideal Poisson-limited experimental measurement. Both quantities use the same Monte Carlo-noise-corrected mask-induced bucket variation, $\sigma_{mask}$, scaled to expected experimental photon counts for $FNR_P$, but normalize it by different noise measures: Monte Carlo statistical uncertainty for $FNR_{MC}$ and photon-counting noise for $FNR_P$.

Having defined the Poisson-limited and Monte Carlo fluctuation-to-noise ratios in equations (3) and (4), respectively, we next use $FNR_P$ to scale the simulated bucket signals to the photon statistics required in a physical acquisition. The Monte Carlo simulation provides, for each illumination pattern i, a simulated bucket signal $B_i$, normalized per primary photon, that is, the mean number of photons crossing the scoring surface per simulated primary photon. These simulated values are then scaled to obtain the expected bucket counts per mAs in a physical acquisition (see Methods). The number of photons transmitted through the mask per realization required to achieve the target $FNR_P$ is:

$$N_{0,req,c} = FNR_{P,target}^2 \left( \frac{\left\langle \sqrt{\sum_j w_{c,j}^2 B_{i,j}} \right\rangle_i}{\sigma_{mask,c}} \right)^2. \quad (5)$$

Here, $FNR_P$=5 and c denotes the detection channel. The quantity $\sigma_{mask,c}$ is the Monte Carlo-noise-corrected standard deviation, $\sigma_{mask}$, for channel c, defined in equation (9). The quantity $B_{i,j}$ is the energy-integrated bucket signal per primary photon at detector j for realization i, and $w_{cj}$ is the weight assigned to detector j in channel c.

Figure 4 summarizes the coupled effects of source spectrum and detector configuration on the bucket-signal strength, mask-induced modulation, Monte Carlo statistical precision, and the resulting ideal photon and dose requirements. The transmission channel retains a high detected-photon yield across all spectra, whereas the scatter yield increases strongly with spectral hardening (Fig. 4a). In contrast, the scatter channels retain substantially larger mask-induced modulation than transmission (Fig. 4b), and their $FNR_{MC}$ increases markedly with photon energy (Fig. 4c). These combined trends reduce the crown dose required for scatter detection at higher photon energies (Fig. 4e), while the required tube loading also depends on the spectrum-specific source output (Fig. 4d).

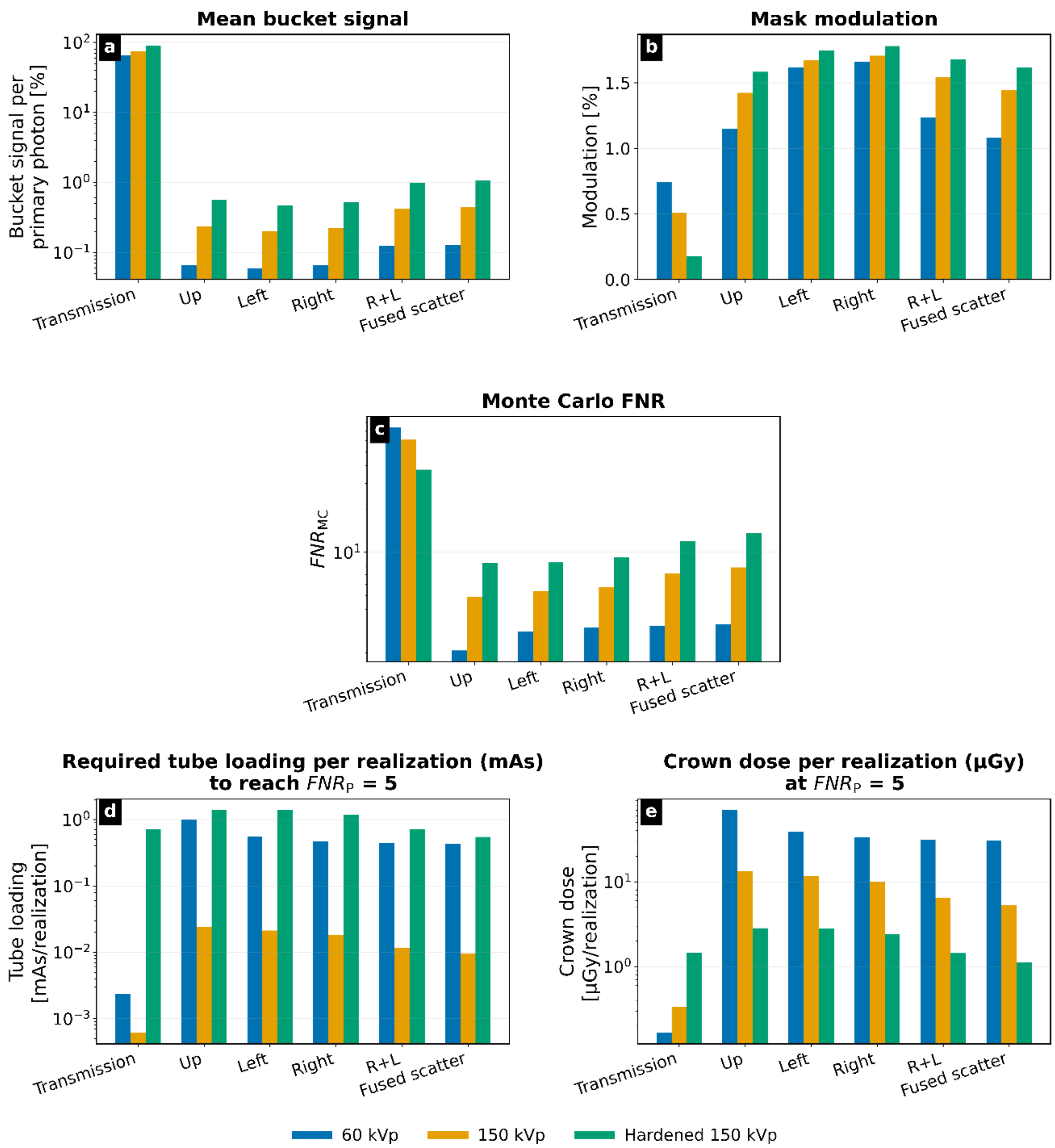

**Fig. 4.** Monte Carlo bucket-signal characteristics and ideal photon-counting feasibility. (a) Mean bucket signal per simulated primary photon, in percent. For the physical detector channels, this corresponds to the mean number of photons crossing the detector surface per simulated primary photon; R+L and fused scatter are weighted signal combinations. (b) Relative mask-induced modulation of the corresponding bucket signals. (c) Monte Carlo fluctuation-to-noise ratio, $FNR_{MC}$, obtained from the simulated bucket statistics. (d) Required tube loading (mAs) per realization to achieve a common bucket $FNR_P = 5$, calculated using the ideal Poisson photon-counting model described in the Methods. (e) Predicted pattern-averaged absorbed dose to the crown required to achieve the same common bucket $FNR_P = 5$. Results are shown for the 60 kVp (blue), 150 kVp (orange), and hardened 150 kVp (green) source spectra. Detector configurations include the transmission channel; the individual Up, Left, and Right scattering channels; the summed lateral scattering signal R+L; and the weighted fused-scatter signal combining the off-axis scattering channels.

As shown in Fig. 4a, the mean detected signal per primary photon increases only modestly for the transmission channel, from 66% at 60 kVp to 90% under hardened 150 kVp, whereas the weighted fused-scatter signal increases by approximately a factor of eight, from 0.13% to 1.06%. Here the signal is the mean number of photons crossing the detector surface per primary photon, where a primary photon is a photon emitted through an open mask pixel. The scored photons include unscattered primary photons as well as scattered and secondary photons, so the value is a ratio of photon numbers rather than the fraction of primaries that reach the detector: 90% means that, for every 10 primary photons, on average 9 photons cross the transmission detector surface. For the combined channels (R+L and fused scatter), the value is a weighted combination of the detector signals, as defined in equation (7). To obtain the corresponding experimental photon signal, this value is multiplied by the number of primary photons transmitted through the mask, as given by equation (6). For example, at 60 kVp, $0.5 \cdot 5.88 \cdot 10^8 = 2.9 \cdot 10^8$ primary photons/mAs pass the mask, giving about $1.9 \cdot 10^8$ photons/mAs at the transmission detector (multiply by 0.66) and about $1.9 \cdot 10^5$ photons/mAs at the Up detector (multiply by 0.065%)[25]. The large difference between the transmission and scatter signals reflects the successive losses along the photon path. At 60 kVp, a transmission signal of 66% means that 34% fewer photons cross the transmission detector than primary photons are emitted. Most photons at those energies that interact with the iron are photo absorbed rather than scattered, each scatter detector collects only part of the solid angle, and scattered photons are further attenuated on their way out of the crown, so that only 0.065% of the primary photons produce a photon crossing the Up detector (see Methods) [18].

The relative mask-induced modulation, calculated from the Monte Carlo-noise-corrected mask-induced standard deviation normalized by the mean bucket signal, as detailed in the Methods, exhibits the opposite spectral trend: transmission-channel modulation decreases from 0.74% to 0.17% with spectral hardening, whereas the scatter channels maintain substantially larger values, ranging from 1.08% to 1.78% (Fig. 4b). The combination of increased scatter-photon yield and sustained mask modulation produces a marked increase in $FNR_{\mathrm{MC}}$ from 2.08–3.14 across the scattering channels at 60 kVp to 8.48–13.52 under hardened 150 kVp (Fig. 4c). Thus, for the harder spectrum, the mask-dependent scattering signal is more clearly resolved relative to the Monte Carlo statistical uncertainty.

A direct consequence of signal-level trends in Fig. 4a-c is the strong reduction in the fluence required for scatter detection under spectral hardening. Assuming Poisson counting statistics, ideal detector efficiency, and no additional noise sources, the relative source fluence required for the five investigated scatter channels to reach the transmission-channel bucket $\mathrm{FNR}_{\mathrm{P}}$ decreases from 183–421 times the transmission requirement at 60 kVp to 15.6–39.7 times at 150 kVp and to 0.76–1.93 times under hardened 150 kVp. When the spectrum-specific source output is taken into account, the corresponding tube loading required to reach $FNR_P = 5$ is 0.43–0.99 mAs per realization for the 60 kVp scatter channels, 0.0096–0.0244 mAs per realization at 150 kVp, and 0.55–1.39 mAs per realization under hardened 150 kVp (Fig. 4d). The high tube-loading requirement of the hardened spectrum reflects the strong reduction in source output produced by the 10-mm Cu filtration. Because the per-pattern object dose scales with source fluence within each spectrum, the same spectral trend is reflected in the dose requirements. The pattern-averaged crown dose per mAs is 71.0, 550 and 2.04 μGy/mAs for the 60 kVp, 150 kVp and hardened 150 kVp spectra, respectively. At the common reference value $FNR_P$=5, the 60 kVp scatter channels require 31–70 μGy per realization, compared with 0.17 μGy for transmission detection. Under hardened 150 kVp, however, the R+L and weighted fused-scatter channels require 1.44 and 1.12 μGy per realization, respectively, compared with 1.47 μGy for the transmission channel (Fig. 4e). Thus, spectral hardening brings the combined scatter channels into the same ideal per-pattern dose range as transmission. For a complete acquisition of 4,000 illumination patterns, each acquired at the target bucket $FNR_P$ of 5, the total tube loading and absorbed dose are 4,000 times the corresponding per-pattern values in Fig. 4d, e.

We next examine whether the signal-level advantage translates into improved reconstruction performance and convergence with the number of mask realizations. Figure 5 presents the CGI reconstructions obtained using all 4,000 illumination patterns. The reference images for the 60 kVp, 150 kVp, and the hardened 150 kVp spectra were obtained with an ideal pixelated flux detector simulated without a mask, as described in the methods. The transmission channels reproduce the crown with high fidelity for all three spectra, whereas the scatter reconstructions depend strongly on detector position and source spectrum. The Up detector recovers the overall crown structure, while the left and right detectors preferentially emphasize the side of the crown facing each detector, reflecting the directional nature of the scattered signal. Combining the scatter channels according to equation (7) yields a more complete representation of the object. Most notably, the quality of the scatter reconstructions improves markedly from 60 kVp to 150 kVp and is highest for the hardened 150 kVp spectrum.

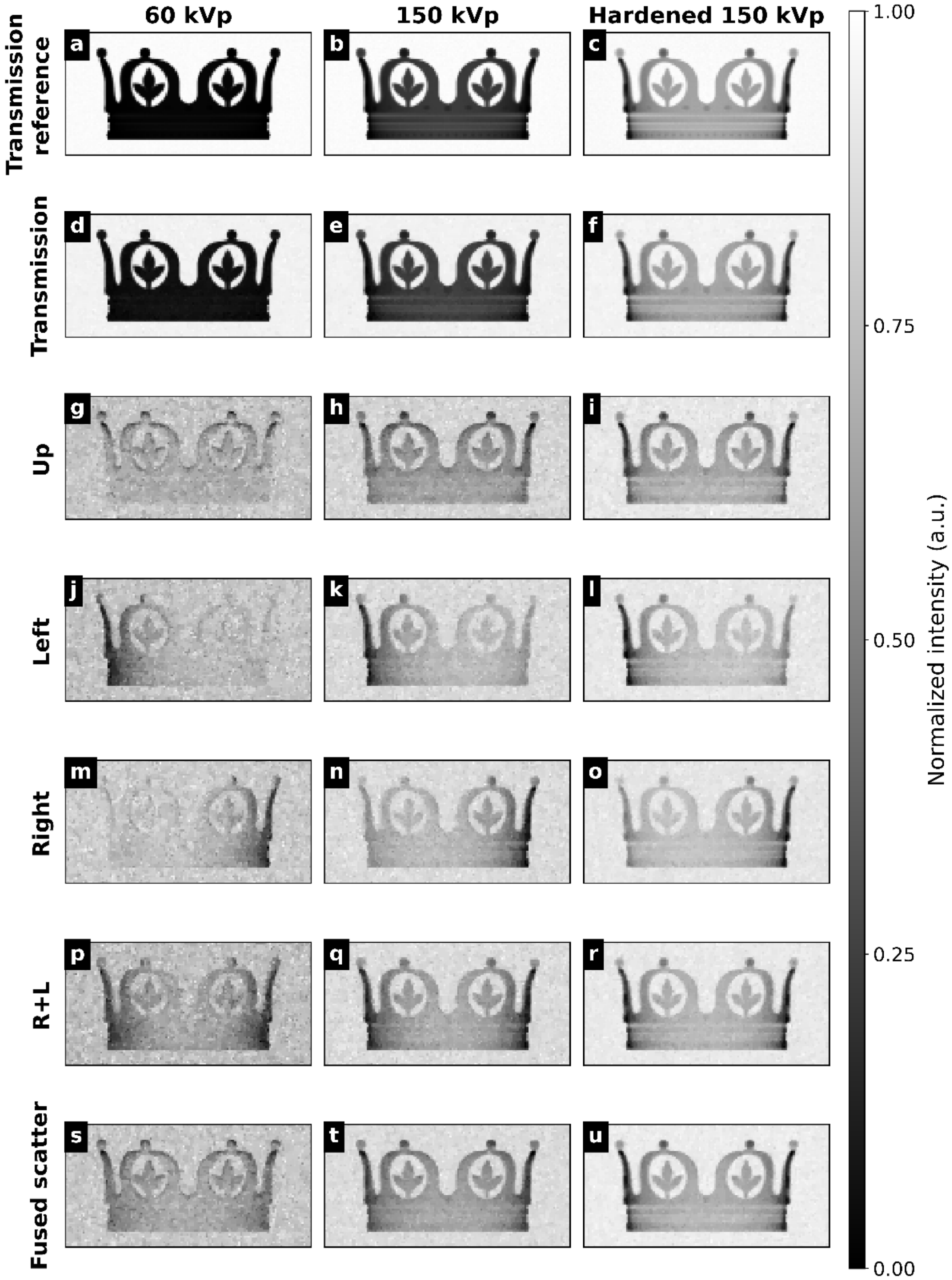

**Fig. 5.** Ghost-image reconstructions obtained from all 4,000 illumination patterns for the three investigated source spectra. The columns correspond to 60 kVp, 150 kVp, and hardened 150 kVp. The rows show, from top to bottom, the transmission reference, transmission reconstruction, Up scatter reconstruction, Left scatter reconstruction, Right scatter reconstruction, the summed lateral scatter reconstruction (R+L), and the weighted fused-scatter reconstruction. All panels are normalized to [0, 1] and share the intensity scale shown on the right. The reconstructions are based on simulated bucket signals containing finite-history Monte Carlo noise ($FNR_{MC}$, Fig. 4c) and therefore do not correspond to a specific experimental dose. The dose per realization required for a bucket $FNR_P$ of 5 is given in Fig. 4e.

To quantify the fidelity of the transmission reconstruction, PSNR and SSIM were calculated relative to the corresponding spectrum-specific mask-free transmission reference (Fig. 6a,b). The transmission reconstruction remained highly consistent across the three investigated spectra, with PSNR values of 28.2–31.2 dB and SSIM values of 0.91–0.96. We next examined convergence with the number of illumination patterns. Because scatter and transmission encode inherently different physical contrast, convergence of each detection channel was evaluated relative to its own reconstruction obtained from the complete 4,000-pattern ensemble, rather than relative to the transmission image. Figure 6c–h therefore shows how rapidly each channel approaches its own full-ensemble reconstruction as the number of realizations increases.

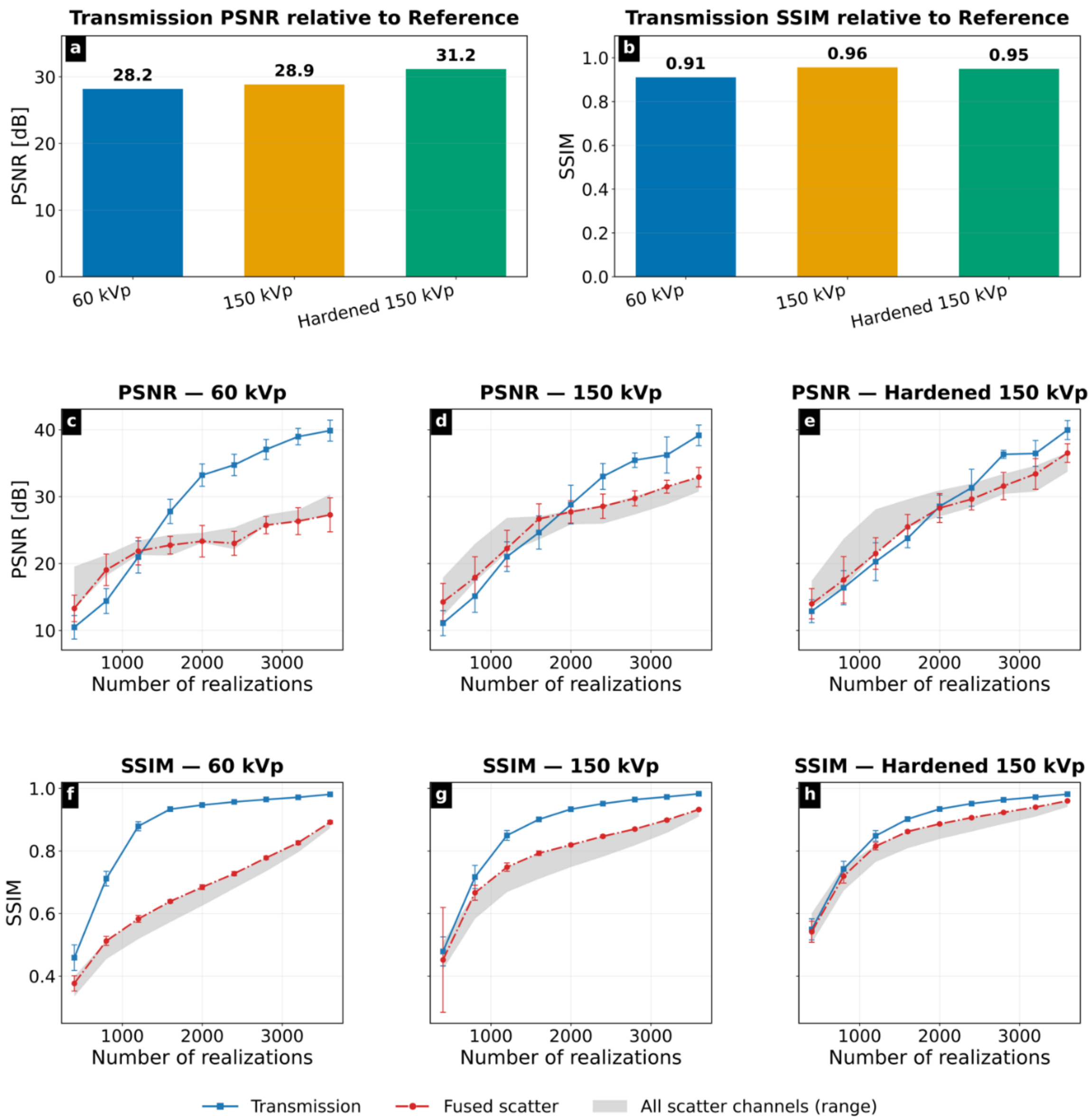

**Fig. 6.** Quantitative transmission fidelity and convergence across source spectra and detection channels. (a,b) Transmission PSNR and SSIM, respectively, for the 4,000-pattern reconstructions relative to the corresponding mask-free transmission reference. (c–h) Self-convergence of PSNR (c–e) and SSIM (f–h) with the number of realizations for 60 kVp (c,f), 150 kVp (d,g), and hardened 150 kVp (e,h). The gray band spans the range of the mean values of all scatter channels (Up, Left, Right, R+L and fused scatter). Points represent the mean over ten random subsets, error bars indicate the corresponding sample standard deviation, and lines connect the sampled means. Transmission is shown in blue and the weighted fused-scatter channel is shown in red as a representative scatter channel. As in Fig. 5, the metrics are computed from the simulated bucket signals and do not correspond to a specific experimental dose.

Figure 6c–h shows that the reconstructions generally approach their corresponding 4,000-pattern reconstructions as the number of illumination patterns increases, with convergence trends that depend on both source spectrum and detection channel. At 60 kVp, the transmission channel

exhibits considerably faster SSIM convergence than the scatter channels; with spectral hardening, the scatter convergence becomes faster and approaches that of transmission (Fig. 6f–h). For example, at 1,200 realizations the SSIM difference between the transmission and fused-scatter channels decreases from 0.30 at 60 kVp to 0.10 at 150 kVp and 0.03 under hardened 150 kVp, consistent with the opposite spectral trends of $FNR_{MC}$ for the two channels (Fig. 4c). The scatter channels (Up, Left, Right and R+L) followed the same convergence trend, with SSIM values within ±13% of the fused-scatter channel over the full range of realizations (gray band). The PSNR and SSIM values quantify agreement with the full-ensemble reconstruction and characterize sensitivity to using fewer illumination patterns. They do not establish reconstruction accuracy or demonstrate that the 4,000-pattern reconstruction has reached an asymptotic limit.

## Discussion

We have presented a Monte Carlo framework for evaluating how source spectrum, detector geometry, and photon-transport physics jointly determine the performance of X-ray CGI. The framework accommodates complex three-dimensional objects, source distributions, and detector arrangements, enabling the prediction of bucket signals and absorbed dose before experimental implementation. Evaluating imaging performance together with radiation exposure is particularly relevant for X-rays because of their ionizing nature. With appropriate modeling of the surrounding materials and shielding, the framework could also support radiation-protection assessments of CGI experiments.

Beyond the specific configuration studied here, the main contribution of this work is in presenting a workflow that turns the design of a ghost imaging system into a physical analysis of the system itself, in two coupled steps: the physical scaling analysis evaluates the expected signal in a certain setup and its modulation due to the masks, whereas the CGI reconstruction analysis evaluates the number of realizations required by a chosen algorithm to reach a specific image quality, and can similarly determine the required signal quality by reconstructing at the corresponding noise level. Linking the two through the number of particles per realization shows where a configuration is limited, and by which component - by the source output, by the mask modulation, or by the reconstruction algorithm, and quantifies the trade-off between image quality, acquisition time and dose before any measurement is made. Such detailed analysis is made possible since the Monte Carlo transport code accounts for the complex interaction physics that governs the signal. This capability is essential for the particle-based GI modalities discussed below, such as neutron, electron and ion ghost imaging and schemes in which the bucket signal is carried by secondary radiation, where the transport physics is considerably more complex than for X-rays and is difficult to capture with simplified forward models. The present workflow therefore provides the tools required for the design and analysis of such advanced systems.

For the iron-crown configuration considered here, our simulations indicate that the transmission reconstruction depends more weakly on the source spectrum than the scatter reconstructions. The scatter reconstructions also exhibit a pronounced dependence on detector position. The Up detector recovers the overall crown structure, whereas the Left and Right detectors preferentially emphasize the side facing each detector. These observations are consistent with the dependence of the bucket signal on the scattering angles, collection solid angle, and photon escape paths associated with each illuminated region. In particular, photons originating from different parts of the crown traverse different material thicknesses before reaching a detector, producing spatial variations in attenuation and sensitivity. Combining the off-axis channels therefore incorporates complementary views of the object and yields a more complete reconstruction without requiring additional object exposures.

The spectral and geometrical dependencies are coupled through photon transport. Changes in photon energy affect both scattering and attenuation, modifying the spatial contributions collected by each detector. For the investigated configuration, spectral hardening increases the scatter-photon yield while maintaining substantial mask-induced modulation, improving the statistical quality of the bucket signals and the stability of the corresponding scatter reconstructions. These trends are consistent with energy-dependent changes in the competition between scattering and absorption. Their individual contributions could be examined further through energy- and interaction-resolved scoring.

The analysis also distinguishes numerical precision, signal-level feasibility, and reconstructed image quality. Monte Carlo uncertainty characterizes the precision of the simulated bucket signals, whereas the separate Poisson model estimates the physical photon budget required to reach a chosen bucket $FNR_P$. The corresponding dose estimates therefore describe an ideal signal-level requirement. Establishing the dose needed for a specified image quality requires reconstruction at the corresponding photon-counting noise level, because equal bucket FNR does not necessarily produce equal PSNR or SSIM. Practical comparisons must also account for the total number of illumination patterns and the reduction in source output caused by spectral filtration, which affects acquisition time.

The interpretation of PSNR and SSIM depends on the selected reference image. Transmission fidelity is evaluated relative to the corresponding spectrum-specific mask-free transmission reference, whereas convergence of each detection channel is evaluated relative to its own reconstruction obtained from the complete 4,000-pattern ensemble. The scatter convergence metrics therefore quantify stability toward the full-ensemble scatter reconstruction and are not interpreted as agreement with the transmission image. This distinction is important because scatter and transmission encode inherently different spatial contrast due to their different photon-transport mechanisms and detector geometries.

These observations motivate the evaluation of reconstruction strategies adapted to off-axis scatter-based CGI. The present inversion recovers an effective two-dimensional response for each detection configuration. Although self-attenuation and angularly dependent transport are included in the simulated bucket signals, their contributions are not separately resolved by the inversion. Joint reconstruction using transport models for the different detectors could relate their complementary measurements to a common description of the object. The improvement obtained through signal-level fusion provides a basis for investigating such approaches, and the Monte Carlo framework can support comparisons of forward models, fusion strategies, and reconstruction algorithms.

The present demonstration assumes ideal binary masks and unit detector efficiency. Extensions could include finite mask contrast, secondary radiation generated in the mask, and energy-dependent detector response, with application-specific benchmarking used to assess predictive accuracy. The quantitative trends reported here are specific to the iron crown, source spectra, and detector arrangement investigated. Other materials and geometries may favor different operating conditions. Within the configurations examined, spectral hardening and complementary detector fusion were favorable for scatter imaging. More generally, the framework provides a means to compare candidate X-ray CGI configurations and guide their selection according to the imaging task, photon budget, and radiation exposure.

Our Monte Carlo transport-based framework is not intrinsically restricted to photons, just as ghost and single-pixel imaging are not restricted to electromagnetic radiation. Recent experiments have demonstrated ghost and single-pixel imaging with thermal and spallation neutrons, relativistic electrons, and correlated ion–electron beams, while related schemes have also been proposed for other massive particles [26–32]. This broader particle-imaging context further motivates a transport-based design framework. Monte Carlo radiation-transport methods can treat energy- and material-dependent interactions, complex geometries, and coupled primary and secondary particle transport within a common numerical framework. This is particularly relevant for modalities in which the detected bucket signal is generated by secondary particles, as recently demonstrated in Monte-Carlo studies of neutron-induced prompt-gamma ghost imaging [33]. The general workflow could therefore be adapted to other ghost imaging modalities using appropriate models of the source, particle transport, and detector response.

## Methods

<u>Monte Carlo system definition</u>

**Monte Carlo implementation.** Monte Carlo simulations were performed using FLUKA version 4-5.1 with Flair version 3.4-5.4. Photon and electron transport thresholds were set to 10 keV. Accordingly, Fe K-shell fluorescence photons (≈6.4–7.1 keV) were not transported and did not contribute to the GI signal; the corresponding energy was deposited locally and included in the absorbed-dose calculation [34]. Each pattern realization comprised $N_p = 5\times10^7$ primary photon histories sampled after application of the binary illumination pattern, i.e., from the open mask pixels, accumulated over five statistically independent FLUKA cycles of $10^7$ primaries each. The same ensemble of 4,000 random binary illumination patterns was used for all three source spectra. For each pattern, every pixel was independently assigned as open with a probability of 0.5.

**Computational performance.** The simulations were performed on a workstation equipped with an Intel® Core™ i7-14700 processor (20 physical cores, 28 processing threads) and 32 GB of RAM, running Ubuntu 24.04 LTS. The mean transport CPU time per pattern realization was $55.68 \pm 6.42$, $73.32 \pm 10.23$, and $99.02 \pm 3.43$ min for the 60 kVp, 150 kVp, and hardened 150 kVp spectra, respectively. Approximately 20–25 independent FLUKA realizations were executed concurrently, and the complete simulation set, comprising 4,000 realizations for each of the three spectra (12,000 realizations in total), required approximately four weeks of wall-clock time.

**Source spectra.** Three polychromatic X-ray spectra were considered: 60 kVp with 2.5 mm Al filtration, 150 kVp with 2.5 mm Al filtration, and a hardened 150 kVp spectrum obtained using 10 mm Cu filtration. All spectra were generated for a tungsten target with a 12° anode angle, using 0.5-keV energy bins, an on-axis geometry at a source-to-plane distance of 100 cm, and a nominal tube loading of 1 mAs. The spectra were generated using SpekPy version 2.0 and converted into normalized cumulative energy distributions for sampling in FLUKA [25]. The structured illumination was implemented as a 4 × 2 cm² planar patterned source, with photons emitted parallel to the optical axis and normal to the mask plane. The patterned source plane coincided with the illumination-mask plane. The corresponding source outputs, $\Phi_S$, over the 8 cm² illumination field were $5.88 \times 10^8$, $4.27 \times 10^9$, and $2.53 \times 10^7$ photons/mAs for the 60 kVp, 150 kVp, and hardened 150 kVp spectra, respectively [25]. For a tube loading q (mAs) per realization, the source emits $q\Phi_s$ photons over the illuminated field, where $\Phi_s$ is the spectrum-specific source output in photons/mAs. The mean number of primary photons transmitted through the binary mask is therefore:

$$N_0(q) = f\, q\, \Phi_s \tag{6}$$

where f is the mean mask open fraction. Each Monte Carlo realization used a fixed number of $N_p = 5 \cdot 10^7$ primary photons sampled after the mask. The experimental scaling was calculated accordingly, using the mean mask open fraction of (f = 0.5).

The spectra shown in Fig. 3b were independently normalized to their respective maximum spectral fluence for visualization only. This normalization was not used for energy sampling or dose conversion. The source illuminated a 4×2 $cm^2$ field represented by 128×64 (P=8192, total number of pixels) pixels with a pitch of 312.5 µm. The mask plane was positioned approximately 1 mm upstream of the crown.

**Illumination mask sampling.** For each primary photon, a custom FLUKA SOURCE routine independently sampled the photon energy from the corresponding cumulative spectrum and uniformly sampled the emission position among the $n_i$ open pixels of illumination pattern i. The simulated bucket signals are therefore normalized per primary photon, that is, per photon transmitted through an open mask pixel. For a tube loading q, the corresponding expected bucket counts are obtained by multiplying the normalized bucket values by $N_0(q)$, as defined in equation (6). For each pattern, a fixed total of $N_p$ primary photons was sampled among its $n_i$ open pixels. The mean number of primary photons sampled per open pixel is therefore $\frac{N_p}{n_i}$. Since the sampling is random, the relative standard deviation in the number of photons sampled per open pixel is approximately $\delta_i \simeq \sqrt{\frac{n_i}{N_p}}$. Accordingly, to limit this relative variation to a specified value $\delta$, $N_p \geq \frac{n_i}{\delta^2}$. For the present masks, $n_i \approx fP = 0.5 \times 8192 = 4096$. A target relative standard deviation of 1% therefore requires $N_p \gtrsim 4.1 \times 10^7$. The selected value $N_p = 5 \times 10^7$ corresponds to approximately $1.2 \times 10^4$ primary photons per open pixel and a relative standard deviation of approximately 0.9%.

**Crown geometry and voxelization.** A three-dimensional crown model in Standard Tessellation Language (STL) format was acquired from the Cults3D online platform and imported into Blender version 4.1 [24,35]. The mesh was edited for voxelization and subsequently exported in OBJ format. A custom MATLAB script was used to import the OBJ file and transform the geometry into a three-dimensional binary voxel matrix using a geometric-processing approach similar to that described by Dong et al. [36]. The resulting matrix was converted into a FLUKA-compatible voxel-geometry file using an adapted version of the FLUKA WRITEGOLEM Fortran program [20]. The resulting voxel model comprised 799×728×400 isotropic voxels, with a voxel side length of 37.57 μm, corresponding to overall dimensions of 30.021×27.353×15.029 $mm^3$. Two region labels were defined in the voxel file, representing the surrounding medium and the crown. The voxelized geometry was incorporated into the simulation using the FLUKA VOXELS card, with iron (density 7.874 $g/cm^3$) assigned to the crown region.

**Bucket detector geometry and position.** The simulated detection geometry comprised one downstream transmission detector and three off-axis scattering detectors, denoted Up, Left, and Right (Fig. 3a). Each detector was represented by a 10 × 10 $cm^2$ ideal planar scoring surface in air. All detector positions were specified relative to the center of the crown's voxel bounding box. The transmission detector was centered 6.0 cm downstream of the crown center, with its scoring surface perpendicular to the incident beam. The Up detector was centered 5.25 cm directly above the crown center, with a horizontal scoring surface. The Right and Left detectors were centered 6.0 and 6.5 cm directly to the right and left of the crown center, respectively, with vertical scoring surfaces parallel to the incident beam.

**Bucket detector scoring.** Each bucket detector was implemented using a USRBDX scorer in one-way current mode. For illumination realization i, the scorer reports the energy-differential photon current $J_i$ crossing the detector scoring surface in the forward direction (2π sr), in units of photons per primary photon per GeV. A single energy bin covering 0–150 keV was used. Since the USRBDX output is expressed per unit energy, the energy-integrated bucket signal for realization (i) was obtained as $B_i=J_i\,\Delta E$, where $\Delta E = 150\text{ keV} = 1.5 \times 10^{-4}\ GeV$. Thus, $B_i$ represents the number of photons crossing the detector scoring surface per primary photon. In addition to the three physical scatter channels, two combined scatter signals were defined as:

$$B_{i,R+L} = B_{i,R} + B_{i,L}, \qquad B_{i,\text{fused}} = B_{i,U} + \frac{1}{2}B_{i,R} + \frac{1}{2}B_{i,L}. \tag{7}$$

**Reference images.** For each source spectrum, a separate mask-free simulation was performed using the same source, crown geometry, and detector configuration, with the full illumination field exposed (no mask). The spatially resolved photon fluence downstream of the crown was scored in air using a photon-fluence USRBIN estimator covering the $4 \times 2\ \text{cm}^2$ transmission field, with the same spatial grid as the illumination patterns - 128×64 pixels with a pitch of 312.5 μm. The scoring volume was 1 cm thick along the beam direction and centered on the transmission scoring plane. The resulting fluence map served as the spectrum-specific reference image for subsequent reconstruction-quality analysis.

**Absorbed dose calculation**. Energy deposition in the crown was scored as dose using a region-based USRBIN estimator assigned to the crown region. Separate open-beam calculations were performed for each source spectrum. The resulting crown dose per mAs was converted to the corresponding pattern-averaged dose per mAs, $D_{pat}$, by accounting for the mean mask open fraction of 0.5.

Reconstruction analysis

**Bucket-signal statistics and modulation.** For each illumination realization $i$, the bucket signal $B_i$ was taken as the processed value obtained from the five statistically independent FLUKA cycles simulated for that pattern. The corresponding absolute Monte Carlo statistical uncertainty, $\sigma_i$, was obtained from the FLUKA-reported relative uncertainty, $\epsilon_i$ , as $\sigma_i = \epsilon_i |B_i|$. Because the observed pattern-to-pattern variation contains contributions from both the mask-dependent signal and finite-history Monte Carlo noise, these contributions were separated before calculating the modulation. The observed standard deviation of the bucket signals was calculated as:

$$\sigma_{obs} = \sqrt{\frac{1}{N-1}\sum_{i=1}^{N}(B_i - \bar{B})^2}\,, \quad \bar{B} = \frac{1}{N}\sum_{i=1}^{N} B_i \tag{8}$$

where N is the number of illumination realizations. The root-mean-square Monte Carlo statistical uncertainty over the realizations was calculated as defined in equation (4). Since the finite-history Monte Carlo fluctuations are statistically independent of the mask-induced signal variation and between realizations, which were simulated with random seeds, the latter was estimated by subtracting the Monte Carlo variance in quadrature:

$$\sigma_{\text{mask}} = \sqrt{\max(\sigma_{\text{obs}}^2 - \sigma_{\text{MC}}^2, 0)} \tag{9}$$

The relative mask-induced modulation was then defined as $M = \frac{\sigma_{mask}}{\bar{B}}$ with the values reported in Fig. 4b expressed as 100M%. The Monte Carlo fluctuation-to-noise ratio, $FNR_{MC}$, was then evaluated using equation (4), thereby quantifying the resolved mask-induced bucket variation relative to the finite-history statistical uncertainty of the Monte Carlo calculation.

**Combination channels.** The scatter signals were combined in the measurement domain before reconstruction so that the complementary scatter signals contributed to a single bucket vector. The fusion weights were selected heuristically based on the spatial complementarity observed among the individual scatter channels: the left and right channels predominantly represent opposite sides of the object, whereas the Up channel provides a more complete response. The two lateral channels were therefore assigned half weight each, giving their combined contribution the same nominal weight as the Up channel. Thus, the relative weights, $w_j$, of the Up, Right, and Left channels were 1:0.5:0.5, as implemented in the combined-channel definitions in equation (7). The same bucket-signal statistics used for the individual detector channels were subsequently evaluated for the combined signals. For each realization i, the absolute Monte Carlo uncertainty of a combined channel c was calculated as $\sigma_{i,c} = \sqrt{\sum_j w_{c,j}^2\, \sigma_{i,j}^2}$, assuming statistically independent detector uncertainties. The corresponding $\sigma_{\mathrm{MC},c}$ was then calculated using equation (4).

**Estimation of experimental photon-counting requirements.** To translate the simulated bucket signals into estimates of the photon fluence and dose required experimentally, we apply a separate ideal Poisson counting model. This analysis is distinct from the Monte Carlo statistical uncertainty described above: the simulated bucket signals represent detector responses per primary photon and are multiplied by the number of photons transmitted through the mask per realization to obtain the expected counts of an ideal photon-counting detector. The ideal counting-statistical $FNR_P$ introduced in the Results (equation (3)) was calculated using the Monte Carlo-noise-corrected mask-induced standard deviation, $\sigma_{\mathrm{mask}}$, defined in equation (9). For a detection channel c, equation (3) was evaluated using $N_0\sigma_{\mathrm{mask},c}$ the bucket-signal fluctuation in photon counts and $\left\langle\sqrt{V_{i,c}}\right\rangle_i$ as the corresponding Poisson noise term, where $N_0$ is the number of photons transmitted through the mask per realization, defined in equation (6), and $V_{i,c}$ is the Poisson counting-noise variance for illumination realization i.

For a weighted combination of statistically independent detector signals,

$$V_{i,c} = \sum_j w_j^2\, s_{i,j} \tag{10}$$

where $w_j$ is the weight assigned to detector $_j$, and $s_{i,j}$ is the expected photon count from detector j for realization i. Accordingly, applying equation (10) to the two combined channels defined in equation (7) gives:

$$V_{i,R+L} = s_{i,R} + s_{i,L}, \quad V_{i,\text{fused}} = s_{i,U} + \frac{1}{4}s_{i,R} + \frac{1}{4}s_{i,L} \tag{11}$$

Since $FNR_P \propto \sqrt{\Phi}$, the relative source fluence required for channel c to reach the transmission-channel $FNR_P$ is calculated as:

$$\frac{\Phi_c}{\Phi_\mathrm{T}} = \left(\frac{FNR_{P,\mathrm{T}}}{FNR_{P,c}}\right)^2 \tag{12}$$

where $\Phi_c$ and $\Phi_\mathrm{T}$ are the source fluences required for channel c and the transmission channel, respectively.

For the absolute feasibility comparison, the common target value of $FNR_P = 5$ introduced in the Results was used. The required number of photons transmitted through the mask per realization was calculated using equation (5), and the corresponding tube loading was obtained using equation (6). The required pattern-averaged crown dose was then calculated as $q_{req,c} = \frac{N_{0,req,c}}{f \cdot \Phi_S}$, $D_{req,c} = q_{req,c} \cdot D_{pat}$ where $D_{pat}$ is the pattern-averaged crown dose per mAs. These estimates assume an ideal 100%-efficient photon-counting detector, statistically independent Poisson counting noise, and no electronic noise or detector-response losses. They therefore represent ideal signal-level feasibility rather than a prediction for a specific physical detector.

**Image reconstruction.** Ghost-images were reconstructed using total-variation (TV) regularization, which promotes sparsity of the image gradient and is well suited to images containing relatively sharp boundaries [37]. The reconstructed image contained 8,192 pixels (128 × 64), whereas 4,000 illumination patterns were used, resulting in an undersampled inverse problem. TV-regularized compressive sensing was therefore used to recover the image from the reduced measurement set. The achievable degree of undersampling depends on the object sparsity, reconstruction method, noise level, and desired reconstruction quality. Because the present work focuses on photon transport and system configuration rather than reconstruction-algorithm optimization, the same 4,000-pattern ensemble was used for all investigated channels and source spectra. The inverse problem defined by equation (1) was solved using the Split-Bregman method [38], as implemented in the PyLops linear-operator and optimization framework [39]. The same reconstruction procedure and parameters were applied to all source spectra and detector channels, without channel-specific optimization.

Before reconstruction, the binary illumination patterns and bucket signals were centered and normalized so that the inversion was driven by realization-to-realization modulation rather than by the absolute source intensity or the mean mask transmission. Because each simulated realization contains a fixed number of primary photons sampled uniformly from the $n_i$ open pixels of pattern i, the expected bucket signal is proportional to $\sum_j W_{ij} x_j$, with $W_{ij} = \frac{M_{ij}}{n_i}$. The sensing matrix was therefore constructed from W rather than from the binary masks, and no row-dependent scaling was applied. Subtracting the mean over realizations from each column of W and from the bucket vector, $B_i - \langle B \rangle = \sum_j \left(W_{ij} - \langle W_j \rangle\right) x_j$, removes the common signal offset associated with the average mask open fraction. The complete reconstruction procedure, including mask and bucket-signal normalization and the numerical parameters used for all reconstructions, is summarized in Table 1.

**Table 1.** TV-regularized ghost-image reconstruction procedure.

| Step | Operation | Implementation |
|---|---|---|
| **Input** | Bucket measurements and binary masks | $B \in \mathbb{R}^N$, $M \in \{0,1\}^{n_x \times n_y \times N}$, $P = n_x \cdot n_y$ |
| **1** | Construct the mask matrix | $M_{ij}, \quad i = 1, \dots, N, \quad j = 1, \dots, P$ |
| **2** | Source weights (fixed number of primaries sampled from the open pixels) | $n_i = \sum_j M_{ij}, \quad W_{ij} = \frac{M_{ij}}{n_i}$ |
| **3** | Center over realizations | $\widetilde{W_{ij}} = W_{ij} - \langle W_j \rangle$ |
| **4** | Common scale and row-count factor | $A_{ij} = s_N P \, \widetilde{W_{ij}}, \quad s_N = \sqrt{\frac{N_{ref}}{N}}, \quad N_{ref} = 4000$ |
| **5** | Normalize the bucket vector | $\sigma_B = \sqrt{\frac{1}{N}\sum_{i=1}^{N}(B_i - \bar{B})^2}$ , $b_i = \frac{s_N(B_i - \bar{B})}{\sigma_B}$ |
| **6** | Construct TV operators | First-order forward finite-difference operators $D_x$ and $D_y$ along the two image dimensions |
| **7** | Solve the TV-regularized inverse problem | $\hat{x} = arg\ min_x \left[\frac{\mu}{2}\|b - Ax\|_2^2 + \varepsilon_{TV}\left(\|D_x x\|_1 + \|D_y x\|_1\right)\right]$, using Split-Bregman optimization |
| **8** | Solver settings | PyLops v2.6.0 Split-Bregman with $\mu = 0.1$, $\epsilon_{TV} = 0.003$ for both TV operators, 25 outer iterations, 2 inner iterations, tolerance $10^{-4}$, and $\tau = 1$. The internal LSQR solver used an iteration limit of 50 and damping $10^{-4}$. |
| **Output** | | Reshape the recovered vector to the image dimensions. |

Both A and b are multiplied by the same factor $s_N$ (Table 1, steps 4–5), so that the data term $\frac{\mu}{2}\|b - Ax\|_2^2$ is scaled by $\frac{N_{ref}}{N}$. Without this factor, the data term would grow in proportion to the number of rows while the TV weight $\epsilon_{TV}$ remains fixed, and reconstructions from more realizations would be effectively less regularized. With this factor, the same μ and $\epsilon_{TV}$ are used for all N, and at $N = N_{ref} = 4000$ the factor equals 1. A solver-sensitivity check was performed by increasing the

LSQR iteration limit from 50 to 100 for all three source spectra and all six reconstruction channels at 400, 1,000, 2,000, 3,000, and 4,000 realizations. The resulting reconstructions were nearly unchanged (SSIM ≥ 0.9978 between the corresponding reconstructions), confirming that the selected LSQR iteration limit did not materially affect the reported convergence behavior.

**Quality metrics.** For each comparison, the reconstructed image and its corresponding reference were transformed to a common physical orientation and independently normalized to the range $[0, 1]$. Image similarity was quantified using peak signal-to-noise ratio (PSNR) and the structural similarity index (SSIM) [40], calculated as

$$PSNR = 10\log_{10}(\frac{L^2}{MSE}) \tag{13}$$

$$SSIM_{(x,y)} = \frac{\left(2\mu_x\mu_y + C_1\right)\left(2\sigma_{x,y} + C_2\right)}{\left(\mu_x^2 + \mu_y^2 + C_1\right)\left(\sigma_x^2 + \sigma_y^2 + C_2\right)} \tag{14}$$

where $L$ is the image dynamic range ($L = 1$ following normalization), MSE is the mean-squared error between the reconstructed and reference images, $\mu_x$ and $\mu_y$ are their mean intensities, $\sigma_x^2$ and $\sigma_y^2$ are the corresponding variances, $\sigma_{xy}$ is their covariance, and $C_1$ and $C_2$ are the standard SSIM stabilization constants [40].

For the final transmission-fidelity analysis, PSNR and SSIM were calculated relative to the corresponding spectrum-specific transmission reference. Scatter-channel fidelity was not quantified relative to the transmission image because the two detection modes encode inherently different physical contrast.

**Convergence**. Convergence was evaluated at 400-realization increments from 400 to 3,600 realizations. At each pattern count, ten subsets were drawn randomly without replacement from the complete 4,000-pattern ensemble. Each subset was independently reconstructed using the procedure summarized in Table 1, and PSNR and SSIM were calculated relative to the corresponding 4,000-pattern reconstruction of the same detection channel and source spectrum. At each pattern count, the reported mean and sample standard deviation summarize the metrics over the ten random subsets. The standard deviation therefore describes variability due to subset selection within the fixed simulated ensemble.

**Use of AI-assisted tools.** ChatGPT (OpenAI) was used to assist with language editing. All AI-assisted edits were reviewed and approved by the authors.

**Code availability.** The custom code used for data analysis and image reconstruction in this study is available from the corresponding author upon reasonable request.

**Data availability**

The datasets generated during and/or analyzed during the current study are available from the corresponding author on reasonable request.

**Funding**

This work was supported by the Israel Science Foundation (Grant No. 2208/24).

**Author contributions**

R.H.S. conceived the study and performed the simulations and data analysis. A.B.-Y., Y. Klein, and O.S. contributed to the development of the GI reconstruction methodology. Y. Korotinsky contributed to the modeling and implementation of the simulation geometry. Y.F. and L.E. contributed to the FLUKA methodology. Y. Klein and Y.F. contributed to the physical interpretation of the results. S. Shwartz supervised the research. R.H.S. and S. Shwartz wrote the manuscript. All authors contributed to the work presented here and to the final paper.

**Competing interests**

The authors declare no competing interests.